\documentclass[preprin,showpacs,preprintnumbers,eqsecnum,amsmath,amssymb,nofootinbib]{revtex4}
\usepackage{graphics,epsfig}
\usepackage{graphicx}
\usepackage{dcolumn}
\usepackage{bm}

\begin{document}
\baselineskip=0.8 cm
\title{ Einstein-Gauss-Bonnet gravity coupled to bumblebee field in four dimensional spacetime}

\author{Chikun Ding$^{1,2,3}$}\thanks{Corresponding author}\email{dingchikun@163.com}
\author{Xiongwen Chen$^{1,3}$}\thanks{Corresponding author}\email{chenxiongwen@hhtc.edu.cn}
\author{Xiangyun Fu$^4$}\email{xyfu@hnust.edu.cn}
\affiliation{$^1$Department of Physics, Huaihua University, Huaihua, 418008, P. R. China\\
$^2$Department of Physics, Hunan University of Humanities, Science and Technology, Loudi, Hunan
417000, P. R. China\\
$^3$Key Laboratory of Low Dimensional
Quantum Structures and Quantum Control of Ministry of Education,
and Synergetic Innovation Center for Quantum Effects and Applications,
Hunan Normal University, Changsha, Hunan 410081, P. R. China\\ $^4$Institute of Physics, Hunan University of Science
and Technology, Xiangtan, Hunan 411201, P. R. China}

\vspace*{0.2cm}
\begin{abstract}
\baselineskip=0.6 cm
\begin{center}
{\bf Abstract}
\end{center}

We study Einstein-Gauss-Bonnet gravity coupled to a bumblebee field which leads to a spontaneous Lorentz symmetry breaking in the gravitational sector. We obtain an exact black hole solution and a cosmological solution in four dimensional spacetime by a regularization scheme. We also obtain a Schwarzschild-like bumblebee black hole solution in $D$-dimensional spacetime.  We find that the bumblebee field doesn't affect the locations of the black hole horizon, but only affects the gravitational potential. That is, its gravitational potential has a minimum value(negative) in the black hole interior  and has a positive value $1+\ell$ at short distance $r\rightarrow0$. If the constant $\ell$ is large enough, then this kind of black hole is practically free from the singularity problem. The thermodynamics and phase transition are also studied. In a cosmological context, it is interesting that the Gauss-Bonnet term has no effect on the conservation of energy equation. A late-time expansion of de Sitter universe can be replicated in an empty space. The Gauss-Bonnet term and the bumblebee field can both actually act as a form of dark energy.

\end{abstract}

 \maketitle

\vspace*{0.2cm}
\section{Introduction}
It is well known that one of the promising theories to unify all interactions is M/string theory \cite{pavluchenko}. In the low energy limit, super-symmetric string theory yields effective field theories in which the Einstein's general relativity is corrected by terms quadratic and higher in the curvature \cite{wiltshire}. The generalization of these corrections is Lovelock's theory \cite{lovelock}, in which General Relativity(GR) is merely the first-order approximation, that cannot describe the effect of strong gravity, such as the interior of a black hole. The only combination of quadratic terms that leads to a ghost-free nontrivial gravitation interaction is the Gauss-Bonnet invariant $\mathcal{G}$ \cite{zwiebach}, which can be seen as the second-order approximation of the gravity theory. So the factor $\alpha$ before $\mathcal{G}$ can be called as an expansion parameter. However, this invariant is a topological invariant in four dimensions, and hence does not contribute to the gravitational dynamics in $D=4$ dimensional spacetime.

Recently, a novel four-dimensional Einstein-Gauss-Bonnet gravity theory was proposed by Glavan and Lin \cite{lin2020} using a regularization scheme. They rescale the expansion parameter $\alpha$ by  $\alpha/(D-4)$ in $D$ dimensions, and then define the four-dimensional theory in the limit $D\rightarrow4$, at the level of the field equations. Some authors pointed out the drawbacks of this regularized scheme \cite{shu,arrechea}. Some authors give several remedies to overcome these objections \cite{hennigar} and obtain the same black hole metric as the original version \cite{lin2020} showing that it is still as a valid solution. Therefore, we can still use the original regularization scheme to study Einstein-Gauss-Bonnet(EGB) gravity coupled to other fields and find some new black hole solutions and cosmological solutions in four dimensional spacetime.  In Ref. \cite{fernandes}, Fernandes obtained the Reissner-Nordstr\"{o}m black hole solution with EGB gravity coupled to Maxwell electric field.  In Ref. \cite{yang} Yang {\it et al} generated the Born-Infeld black holes with EGB gravity minimally coupled to the Born-Infeld nonlinear electromagnetic field. In Ref. \cite{feng}, Feng {\it et al} studied the cosmological solution with EGB gravity coupled to a scalar field. The Bardeen-like and Hayward-like black holes were derived in Refs. \cite{kumar2004}. In Refs. \cite{wei2021}, Wei {\it et al} investigated the rotating black hole solutions. Some other black hole solutions were discussed in the Refs. \cite{ghosh}.

The string theory also involves considering a spontaneous Lorentz symmetry breaking \cite{kostelecky198939} due to that Lorentz invariance(LI) should not be a truth at very high energy scales  \cite{mattingly}. Other proposals of Lorentz violation(LV) include noncommutative field theories \cite{carroll}, spacetime-varying fields \cite{bertolami69}, loop quantum gravity theory \cite{gambini}, brane world scenarios \cite{burgess03}, massive gravity \cite{fernando} and Einstein-aether theory \cite{jacobson}, etc. At experimentally accessible scales, the possible LV observable signals can be described by an effectively field theory---the standard model extension(SME) \cite{kostelecky2004}.

GR and the standard model(SM) in particle physics are both based on LI and the spacetime backdrop, but they process two different energy scales, i.e., low energy for GR, high energy for SM. However, at very high energy scale (Planck energy scale), the GR effect of a particle cannot ignorable. Hence an effective field theory known as SME is set up to couple the SM to GR. It introduces an extra things which may includes information about the LV at the Plank scale \cite{kostelecky2004}. The straightforward item is the bumblebee field, a single vector $B_\mu$. It has a nonzero vacuum expectation value and its smooth quadratic potential unfolds the spontaneous LV. Kostelecky and Samuel first studied this bumblebee\footnote{The name ``bumblebee" of this vector field is given by V. A. Kosteleck\'{y} \cite{bluhm125007}.} gravitational model  in 1989 \cite{kostelecky198939,kostelecky198940,dickinson} as a specific way of active Lorentz violation.

So in this paper, we would like to study this effective field theory---Einstein-Gauss-Bonnet gravity coupled to the bumblebee fields, to get some suppressed effects emerging from the underlying unified quantum gravity theory, on our low energy scale.  In 2018, R. Casana {\it et al} obtained a Schwarzschild-like black hole solution in the bumblebee gravity model and researched its some classical detections \cite{casana}.  The rotating black hole solutions are the most relational subsets for astrophysics. In 2020, we found an exact Kerr-like solution through solving Einstein-bumblebee gravitational field equations and studied its black hole shadow\cite{ding2020}.  Then Li and \"{O}vg\"{u}n \cite{li} study the weak gravitational deflection angle of relativistic massive particles by this Kerr-like black hole. Jha and Rahaman \cite{jha} extended this Kerr-like solution to Kerr-sen case.

We will study  the black hole solutions and cosmological solutions in the theory of Einstein-Gauss-Bonnet gravity coupled to the bumblebee fields in four dimensional spacetime. The rest paper is organized as follows.   In Sec. II we give the background for the Einstein-Gauss-Bonnet-bumblebee theory. In Sec. III, we give the  black hole solution by solving the gravitational field equations. In Sec. IV, we study its cosmological solutions and find some effects of the Lorentz breaking constant $\ell$. Sec. V is for a summary.

\section{Einstein-Gauss-Bonnet-bumblebee theory}

Since EGB gravity is the second-order approximation of the gravity theory and can describe the effects of the strong gravity, it is natural to consider SM coupling to EGB gravity, so one can find some LV effect in the strong gravity sector. In the bumblebee gravity model, one introduces the bumblebee vector field $B_{\mu}$ which has a nonzero vacuum expectation value, to lead a spontaneous Lorentz symmetry breaking in the gravitational sector via a given potential. In $D-$dimensional spacetime, the
 action of Einstein-Gauss-Bonnet gravity coupled to this bumblebee field is \cite{ding2021},
\begin{eqnarray}
\mathcal{S}=
\int d^Dx\sqrt{-g}\Big[\frac{R}{2\kappa}+\frac{2\alpha}{D-4} \mathcal{G}+\frac{\varrho}{2\kappa} B^{\mu}B^{\nu}R_{\mu\nu}-\frac{1}{4}B^{\mu\nu}B_{\mu\nu}
-V(B_\mu B^{\mu}\mp b^2)+\mathcal{L}_M\Big], \label{action}
\end{eqnarray}
where $R$ is Ricci scalar, $\kappa=2G\int d\Omega_{D-2}$ which is the production between the Newton constant $G$ and the area of a unit $D-2$ sphere.
$\alpha$ is an expansion parameter and the Gauss-Bonnet invariant $\mathcal{G}$ is,
\begin{eqnarray}\label{}
\mathcal{G}=R_{\mu\nu\tau\sigma}R^{\mu\nu\tau\sigma}-4R_{\mu\nu}R^{\mu\nu}+R^2,
\end{eqnarray}
which can be seen as the second-order approximation of the gravity theory and may sufficiently describe the strong gravity effects.
In four dimensions, it is a topological invariant and has no contribution to the gravitational dynamics. But if one uses a regularization scheme---re-scaling $\alpha$ by $\alpha/(D-4)$ and taking limit $D\rightarrow4$, Glavan and Lin \cite{lin2020} found  this Gauss-Bonnet term can make a non-trivial effect.

The coupling constant $\varrho$ dominates the non-minimal gravity interaction to bumblebee field $B_\mu$. The term $\mathcal{L}_M$ represents possible interactions with matter or external currents.
The constant $b$ is a real positive constant. The potential $V(B_\mu B^{\mu}\mp b^2)$ triggers Lorentz and/or $CPT$ (charge, parity and time) violation. It gives a nonzero vacuum expectation value (VEV) for bumblebee field $B_{\mu}$ indicating that the vacuum of this model obtains a prior direction in the spacetime. This potential has a minimum at $B^{\mu}B_{\mu}\pm b^2=0$ and $V'(b_{\mu}b^{\mu})=0$ to ensure the destroying the $U(1)$ symmetry, where the field $B_{\mu}$ acquires a nonzero VEV, $\langle B^{\mu}\rangle= b^{\mu}$. Another vector $b^{\mu}$ is a function of the spacetime coordinates and has a constant value $b_{\mu}b^{\mu}=\mp b^2$, where $\pm$ signs mean that $b^{\mu}$ is timelike or spacelike, respectively.
The bumblebee field strength is
\begin{eqnarray}
B_{\mu\nu}=\partial_{\mu}B_{\nu}-\partial_{\nu}B_{\mu}.
\end{eqnarray}
This antisymmetry of $B_{\mu\nu}$ implies the constraint \cite{bluhm}
\begin{eqnarray}
\nabla ^\mu\nabla^\nu B_{\mu\nu}=0,
\end{eqnarray} which means that the conservation law holds even though the potential $V$ excludes $U(1)$ gauge symmetry. In Minkowski and Riemann spacetime, this model provides a dynamical theory generating a photon as a Nambu-Goldstone boson for spontaneous Lorentz violation \cite{bluhm2005} and without generating ghosts.

Varying the action (\ref{action}) with respect to the metric yields the gravitational field equations
\begin{eqnarray}\label{einstein0}
G_{\mu\nu}=R_{\mu\nu}-\frac{1}{2}g_{\mu\nu}R=\kappa T_{\mu\nu}^B+2\alpha\kappa T^{GB}_{\mu\nu}+\kappa T_{\mu\nu}^M,
\end{eqnarray}
where the bumblebee energy momentum tensor $T_{\mu\nu}^B$ is
\begin{eqnarray}\label{momentum}
&&T_{\mu\nu}^B=B_{\mu\alpha}B^{\alpha}_{\;\nu}-\frac{1}{4}g_{\mu\nu} B^{\alpha\beta}B_{\alpha\beta}- g_{\mu\nu}V+
2B_{\mu}B_{\nu}V'\nonumber\\
&&+\frac{\varrho}{\kappa}\Big[\frac{1}{2}g_{\mu\nu}B^{\alpha}B^{\beta}R_{\alpha\beta}
-B_{\mu}B^{\alpha}R_{\alpha\nu}-B_{\nu}B^{\alpha}R_{\alpha\mu}\nonumber\\
&&+\frac{1}{2}\nabla_{\alpha}\nabla_{\mu}(B^{\alpha}B_{\nu})
+\frac{1}{2}\nabla_{\alpha}\nabla_{\nu}(B^{\alpha}B_{\mu})
-\frac{1}{2}\nabla^2(B^{\mu}B_{\nu})-\frac{1}{2}
g_{\mu\nu}\nabla_{\alpha}\nabla_{\beta}(B^{\alpha}B^{\beta})\Big],
\end{eqnarray}
and the Gauss-Bonnet energy momentum tensor $T^{GB}_{\mu\nu}$ is,
\begin{eqnarray}\label{momentumG}
&&T_{\mu\nu}^{GB}=4R_{\alpha\beta}R^{\alpha\;\beta}_{\;\mu\;\nu}-2R_{\mu\alpha\beta\gamma}
R_{\nu}^{\;\alpha\beta\gamma}+4R_{\mu\alpha}R^{\alpha}_{\;\nu}-2RR_{\mu\nu}
+\frac{1}{2}g_{\mu\nu}\mathcal{G}.
\end{eqnarray}
The prime denotes differentiation with respect to the argument,
\begin{eqnarray}
V'=\frac{\partial V(x)}{\partial x}\Big|_{x=B^{\mu}B_{\mu}\pm b^2}.
\end{eqnarray}
Varying instead with respect to the the bumblebee field generates the bumblebee equations of motion (supposing that there is no coupling between the bumblebee field and $\mathcal{L}_M$),
\begin{eqnarray}\label{motion}
\nabla ^{\mu}B_{\mu\nu}=2V'B_\nu-\frac{\varrho}{\kappa}B^{\mu}R_{\mu\nu}.
\end{eqnarray}

The contracted Bianchi identities ($\nabla ^\mu G_{\mu\nu}=0$) lead to conservation of the total energy-momentum tensor
\begin{eqnarray}\label{}
\nabla ^\mu T_{\mu\nu}=\nabla ^\mu\big( T^B_{\mu\nu}+2\alpha T^{GB}_{\mu\nu}+T^M_{\mu\nu}\big)=0.
\end{eqnarray}

In the next sections, we derive the  black hole solution and cosmological solution by solving  gravitational equations in this Einstein-Gauss-Bonnet-bumblebee model.

\section{Black hole solution in Einstein-Gauss-Bonnet-bumblebee gravity}
In this section, we suppose that there is no matter field and the bumblebee field is frosted at its VEV like in Refs \cite{casana,bertolami}, i.e., it is
\begin{eqnarray}
B_\mu=b_\mu,
\end{eqnarray}
then the specific form of the potential controlling its dynamics is irrelevant.
And  as a result, we have $V=0,\;V'=0$. Then the first two terms in Eq. (\ref{momentum}) are like those of the electromagnetic field, the only distinctive are the coupling items to Ricci tensor. Under this condition,  Eq. (\ref{einstein0}) leads to gravitational field equations \cite{ding2021}
\begin{eqnarray}\label{bar}
G_{\mu\nu}=2\alpha\kappa T^{GB}_{\mu\nu}+\kappa (b_{\mu\alpha}b^{\alpha}_{\;\nu}-\frac{1}{4}g_{\mu\nu} b^{\alpha\beta}b_{\alpha\beta})+\varrho\Big(\frac{1}{2}
g_{\mu\nu}b^{\alpha}b^{\beta}R_{\alpha\beta}- b_{\mu}b^{\alpha}R_{\alpha\nu}
-b_{\nu}b^{\alpha}R_{\alpha\mu}\Big)
+\bar B_{\mu\nu},
\end{eqnarray}
with
\begin{eqnarray}\label{barb}
&&\bar B_{\mu\nu}=\frac{\varrho}{2}\Big[
\nabla_{\alpha}\nabla_{\mu}(b^{\alpha}b_{\nu})
+\nabla_{\alpha}\nabla_{\nu}(b^{\alpha}b_{\mu})
-\nabla^2(b_{\mu}b_{\nu})-g_{\mu\nu}\nabla_\alpha\nabla_\beta(b^\alpha b^\beta)\Big].
\end{eqnarray}
The static spherically symmetric black hole metric in a $D$ dimensional spacetime  have the general form
\begin{eqnarray}\label{metric}
&&ds^2=-e^{2\phi(r)}dt^2+e^{2\psi(r)}dr^2+r^2d\Omega_{D-2}^2,
\end{eqnarray}
where $\Omega_{D-2}$ is a standard $D-2$ sphere.

In the present study, we pay attention to that the bumblebee field has a radial vacuum energy expectation because that the spacetime curvature has a strong radial variation, on the contrary that the temporal changes are very slow. So the bumblebee field is supposed to be spacelike($b_\mu b^\mu=$ positive constant) as that
\begin{eqnarray}\label{bu}
b_\mu=\big(0,be^{\psi(r)},0,0,\cdots,0\big),
\end{eqnarray}
where $b$ is a positive constant.
Then the bumblebee field strength is
\begin{eqnarray}
b_{\mu\nu}=\partial_{\mu}b_{\nu}-\partial_{\nu}b_{\mu},
\end{eqnarray}
whose components are all zero. And their divergences are all zero, i.e.,
\begin{eqnarray}
\nabla^{\mu}b_{\mu\nu}=0.
\end{eqnarray}
From the equation of motion (\ref{motion}), we have
\begin{eqnarray}
b^{\mu}R_{\mu\nu}=0\label{motion2}.
\end{eqnarray}
The gravitational field equations (\ref{bar}) become
\begin{eqnarray}\label{}
G_{\mu\nu}=2\alpha\kappa T^{GB}_{\mu\nu}+\bar B_{\mu\nu}.
\end{eqnarray}

 For the metric (\ref{metric}), the nonzero components of Einstein tensor $G_{\mu\nu}$, the Gauss-Bonnet momentum tensor $T^{GB}_{\mu\nu}$ and the bumblebee tensor $\bar B_{\mu\nu}$ are shown in the appendix.
 By using the motion equation (\ref{motion2})
\begin{eqnarray}R_{11}=\frac{(D-2)}{r}\psi'-(\phi''+\phi'^2-\phi'\psi')=0,\end{eqnarray}
one can obtain the following three gravitational field equations
\begin{eqnarray}
&&(D-3)(e^{2\psi}-1)+2r\psi'=-4(D-3)\alpha\kappa\Big[\frac{2}{r}(1-e^{-2\psi})\psi'
+\frac{D-5}{r^2}\Big(\frac{e^{2\psi}+e^{-2\psi}}{2}-1\Big)\Big]\nonumber\\
&&\qquad\qquad\qquad\qquad\qquad\qquad+\ell\big[(D-3)-2r\psi'\big],\label{tt}\\
&&(D-3)(1-e^{2\psi})+2r\phi'=4(D-3)\alpha\kappa\Big[-\frac{2}{r}(1-e^{-2\psi})\phi'
+\frac{D-5}{r^2}\Big(\frac{e^{2\psi}+e^{-2\psi}}{2}-1\Big)\Big]\nonumber\\
&&\qquad\qquad\qquad\qquad\qquad\qquad-\ell\big[(D-3)+2r\phi'\big],\label{rr}\\
&&\frac{(D-3)(D-4)}{2}(1-e^{2\psi})+r\psi'+(D-3)r\phi'=2(D-3)\alpha\kappa
\Big\{-4e^{-2\psi}\phi'\psi'+(1-e^{-2\psi})\Big[
\frac{(D-5)(D-6)}{2r^2}\nonumber\\
&&\qquad\qquad\qquad\qquad\cdot(e^{2\psi}-1)-\frac{2(D-5)}{r}\phi'-\frac{6}{r}\psi'\Big]\Big\}
-\ell\Big[\frac{(D-3)(D-4)}{2}+r\psi'+(D-3)r\phi')\Big]\label{theta},
\end{eqnarray}
where we have redefined the Lorentz-violating parameter $\ell=\varrho b^2$ and, the prime $'$ is the derivative with respect to the corresponding argument, respectively.

Adding the Eq. (\ref{tt}) to the  Eq. (\ref{rr}), one can obtain the result $\phi'(r)+\psi'(r)=0$, since the alternative solution $1+2\ell+2(D-3)\alpha\kappa(1-e^{-2\psi})/r^2=0$ is incompatible with the remaining field equations.
Rewriting the function $e^{2\psi}=1/f(r)$, then one can change the Eq. (\ref{tt}) as the form
\begin{eqnarray}
(D-3)[r^2(1-f-\ell f)+2\alpha\kappa(D-5)(1-f)^2]=rf'[(1+\ell)r^2+4(D-3)\alpha\kappa(1-f)].
\end{eqnarray}
It can be written
\begin{eqnarray}
\frac{d}{dr}\big[r^{D-3}(1-f-\ell f)+2(D-3)\alpha\kappa r^{D-5}(1-f)^2\big]=0.
\end{eqnarray}
Its solution is
\begin{eqnarray}\label{dmetric}
f(r)=1+\frac{(1+\ell)r^2}{2E}\left[1\pm\sqrt{1+\frac{4E\ell}{(1+\ell)^2r^2}
+\frac{8EGM}{(1+\ell)^2r^{D-1}}}\right],
\end{eqnarray}
where $E=2(D-3)\alpha\kappa$, $M$ is the mass of the black hole. Note that if $\ell=0$ and $E=2(D-3)(D-4)\alpha\kappa$, it is the same as that in \cite{boulware}.
Setting $e^{2\phi}=A(r)f(r)$ and using the result $\psi'+\phi'=0$, one can obtain $A(r)$ is an arbitrary constant. Here we can let this arbitrary constant to $A=1+\ell$ for convenience, so,
\begin{eqnarray}\label{dell}
e^{2\phi(r)}=(1+\ell)f(r).
\end{eqnarray}

When $D\rightarrow4$, the black hole metric in the bumblebee gravity is
\begin{eqnarray}\label{bmetric}
ds^2=- (1+\ell)f(r)dt^2+\frac{1}{f(r)}dr^2+r^2d\theta_1^2
+r^2\sin^2\theta_1 d\theta_2^2,
\end{eqnarray}with
\begin{eqnarray}\label{metricf}
f(r)=1+\frac{(1+\ell)r^2}{32\pi\alpha G}\left[1\mp\sqrt{1+\frac{64\pi\alpha G\ell}{(1+\ell)^2r^2}
+\frac{128\pi\alpha G^2M}{(1+\ell)^2r^3}}\right].
\end{eqnarray}
There are two branches of solutions for the metric function (\ref{metricf}) with $\alpha>0$, the first one corresponding to $``-"$ sign; the second to $``+"$ sign.
Lastly, substituting these quantities into Eqs. (\ref{metric}) and (\ref{bu}), we can get the bumblebee field $b_\mu=(0,1/\sqrt{f(r)},0,0)$.

If $\ell\rightarrow0$, it recovers the four dimensional Gauss-Bonnet black hole metric \cite{lin2020}.
When $\alpha\rightarrow0$, the first branch becomes
\begin{eqnarray}\label{}
ds^2=- \Big(1-\frac{2GM}{r}\Big)dt^2+\frac{1+\ell}{1-2GM/r}dr^2+r^2d\theta_1^2
+r^2\sin^2\theta_1 d\theta_2^2,
\end{eqnarray}
a Schwarzschild-like black hole which is the same as that in Ref. \cite{casana}; the second one becomes
\begin{eqnarray}\label{desitter}
f(r)=\frac{1}{1+\ell}\Big[
1+2\ell+\Lambda r^2+\frac{2GM}{r}\Big],
\end{eqnarray}
which is a Schwarzschild-de Sitter like black hole with a negative mass $M$ and a positive cosmological constant $\Lambda=(1+\ell)^2/16\pi\alpha G$.  It has been shown by Boulware and Deser \cite{boulware} that the second branch is unstable and leads to graviton ghost.

When $r\rightarrow\infty$, the two branches have similar behaves asymptotically at large distances, i.e., the first is asymptotically to a Schwarzschild-like black hole with positive mass $M$, the second to a Schwarzschild-de Sitter like black hole with negative mass $M$.

Now we focus on the first solution, i.e., the minus sign ``$-$" of the metric (\ref{metricf}) and let $16\pi G=1$, that is,
\begin{eqnarray}\label{metricm}
f(r)=1+\frac{(1+\ell)r^2}{2\alpha }\left[1-\sqrt{1+\frac{4\alpha \ell}{(1+\ell)^2r^2}
+\frac{8\alpha M}{(1+\ell)^2r^3}}\right].
\end{eqnarray}
 It is easy to see that their two horizon, locates at $g_{00}(r_\pm)=0$,
\begin{eqnarray}\label{}
r_\pm^H=M\pm\sqrt{M^2-\alpha}.
\end{eqnarray}
which are the same as those in Ref. \cite{lin2020}.
The horizon at $r_+^H$ is the event horizon of a black hole, and $r_-^H$ is the event horizon of a white black hole. If the mass $M$ is smaller than the critical mass which is given by
\begin{eqnarray}\label{}
M_*=\sqrt{\alpha},
\end{eqnarray}
there is no horizon forming and hence no black hole solutions.

Note that the black hole horizon doesn't depend on $\ell$, i.e., the bumblebee field doesn't affect the location of black hole horizon.
 \subsection{Free from singularity problem?}
In the Fig. 1, we plot the radial dependence of $-g_{00}$ with different $\alpha$\footnote{$\alpha$ possesses a dimension of length$^2$ in 4 dimensions, so its values in the figure 1 to 3 are defined as $\alpha\rightarrow \alpha/M^2$.} and $\ell$. One can see that the gravitational potential has only one minimum, and approaches a finite value $1+\ell$ at short distances $r\rightarrow0$, which is different from the Schwarzschild geometry. This minimum locates at $r_m$ from $f'(r_m)=0$,
\begin{eqnarray}\label{}
(1+\ell)^2Mr_m^3-\alpha(\ell r_m+M)^2=0.
\end{eqnarray}
\begin{figure}[ht]\label{g00}
\begin{center}
\includegraphics[width=5.0cm]{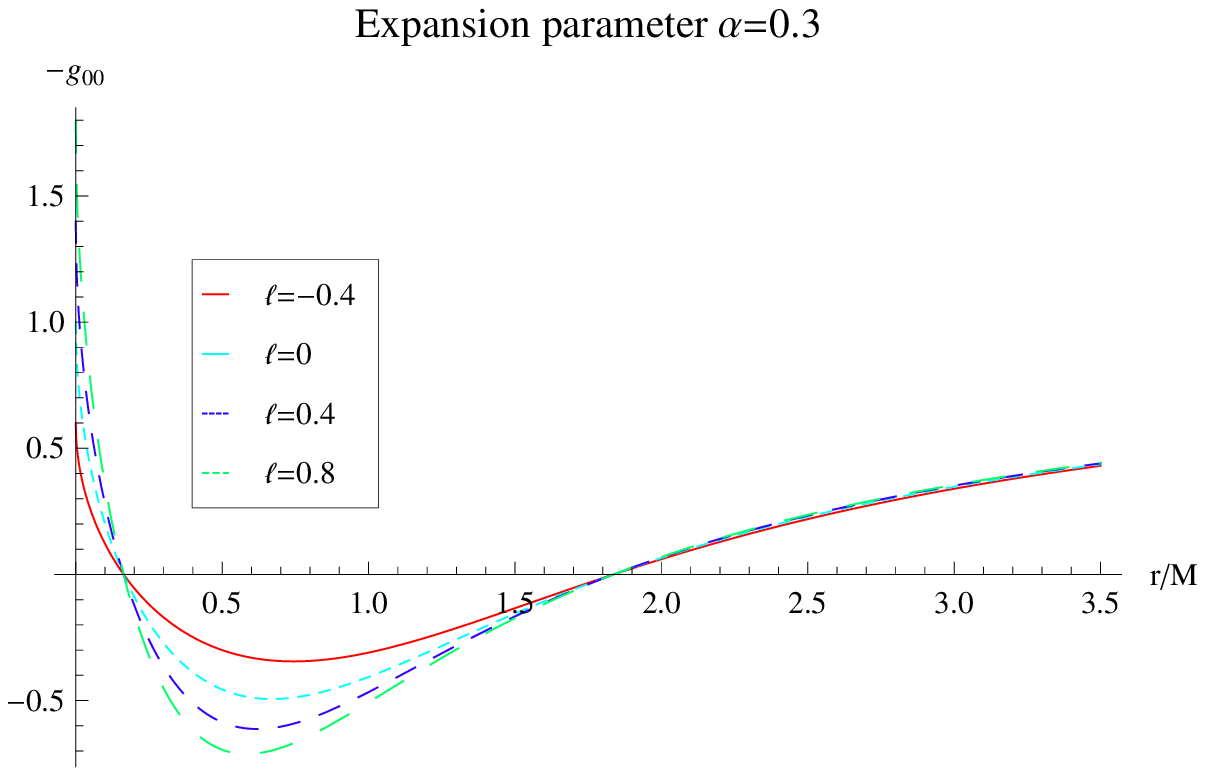}\;\;\;\;\includegraphics[width=5.0cm]{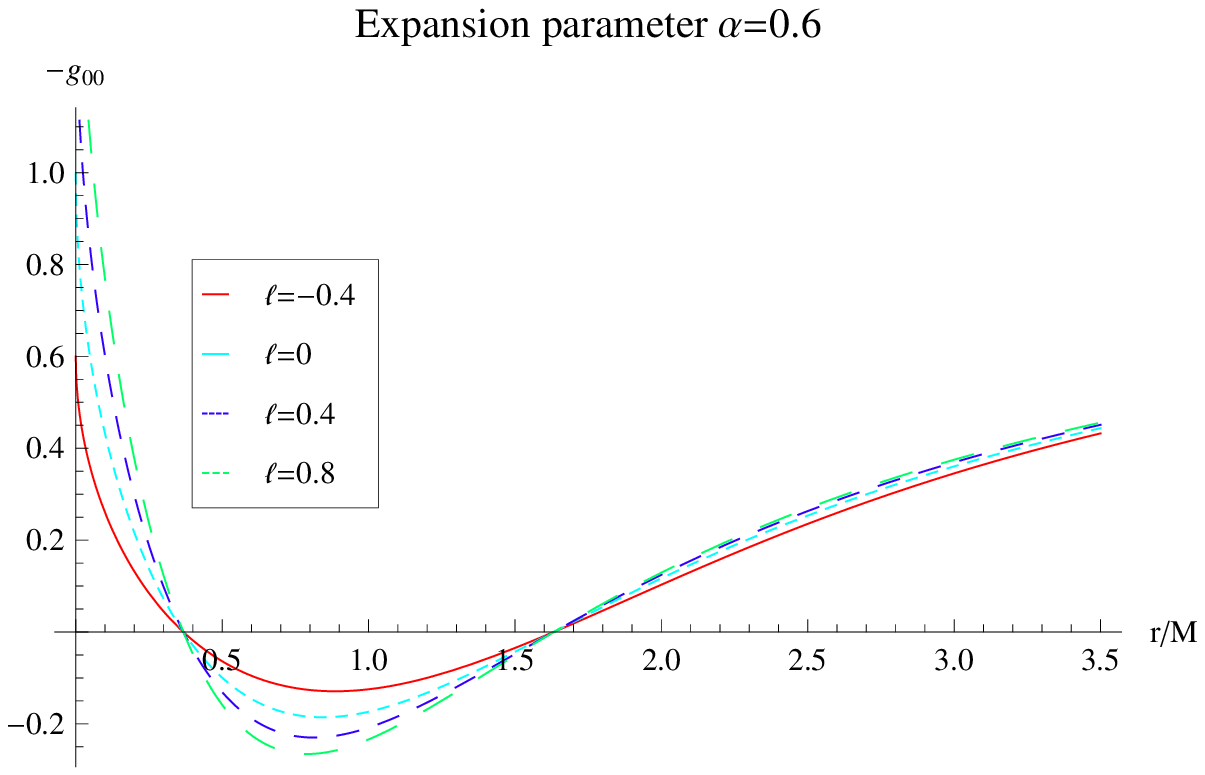}
\;\;\;\;\includegraphics[width=5.0cm]{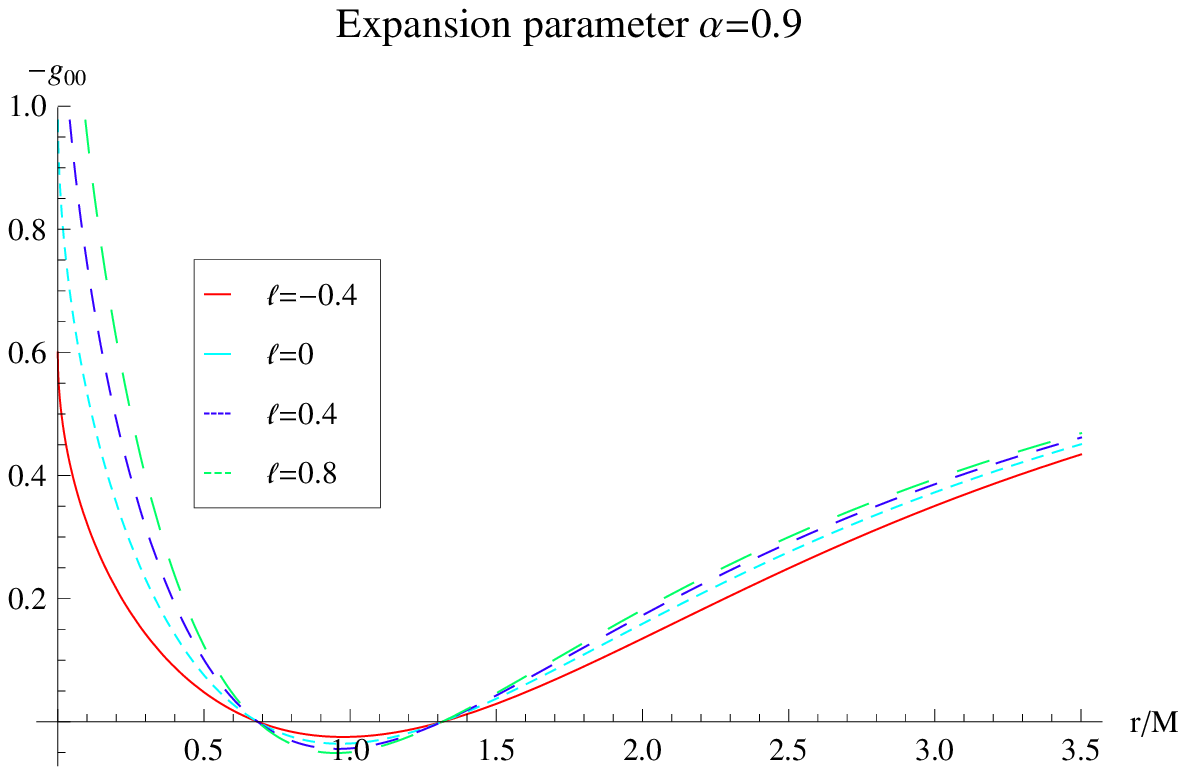}
\caption{Radial dependence of gravitational potential $-g_{00}$ with different coupling constant $\ell$ in the case of different expansion parameter $\alpha$. Note that $\alpha=0.3,0.6,0.9$ correspond to the black hole mass $M=M_*/\sqrt{0.3}, M=M_*/\sqrt{0.6}, M=M_*/\sqrt{0.9}$, respectively. }
\end{center}
\end{figure}
From the Fig. 1, one can see that the LV constant $\ell$ pushes $r_m$ to the origin $r=0$.

In Refs. \cite{lin2020,yang}, the gravitational potential $-g_{00}$ approaches unit $-g_{00}\rightarrow1$. So an infalling particle would feel a repulsive force and seems never reach the origin $r\rightarrow0$, and may be free from the singularity problem. In Ref. \cite{yang}, Yang {\it et al} found that, if the infalling particle starts at rest at a finite distance $R$, it indeed can't reach the singularity. However, if it starts at rest at infinity, then it can just reach the singularity with zero speed \cite{arrechea}; if the particle has a kinetic energy at infinity, it will reach the singularity with a nonzero speed \cite{yang} and the black hole solution still suffers from the singularity problem. But in the case of coupling to the bumblebee field, the gravitational potential $-g_{00}$ has a bigger value $\ell+1$ than unit at $r\rightarrow0$. So one can say the picture is different.

 When the particle starts at rest at radius $R$ and freely falls radially to the black hole, the velocity  reads \cite{chandrasekhar},
\begin{eqnarray}\label{}
v(r)=\frac{dr}{d\tau}=\pm\sqrt{f(R)-f(r)},
\end{eqnarray}
the particle's energy $E$ per unit rest mass is $E^2=(1+\ell)f(R)$, and plus (minus) sign refers to the infalling (outgoing) particle. Fig. 3 shows the behavior of velocity of an infalling particle which starts at rest at different finite distance $R$ (Left) and at a finite distance $R$ with different $\ell$ (Right). This particle then accelerates to a maximum velocity under the attractive force, nextly decelerates to zero speed at a short distance by the repulsive force. After that, it begins to go out and back to its initial point. So that it cannot reach the origin $r=0$.
\begin{figure}[ht]\label{}
\begin{center}
\includegraphics[width=7.0cm]{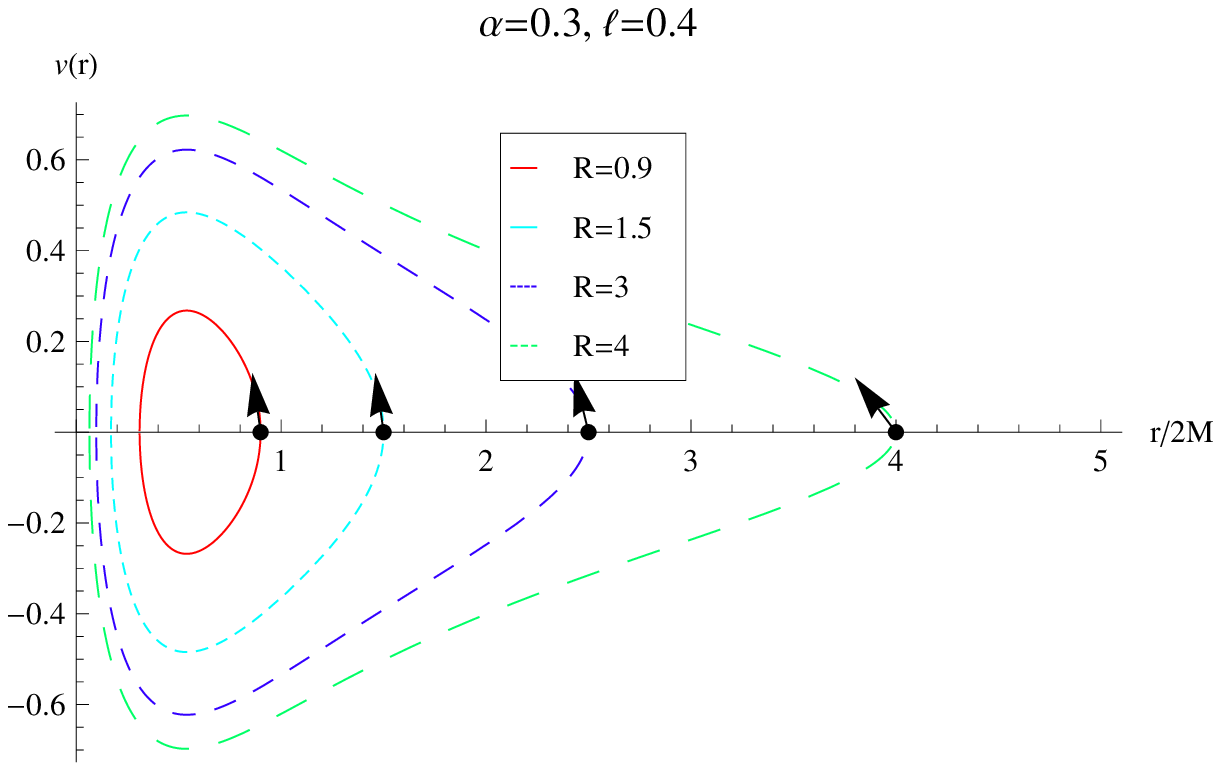}\;\;\;\;\includegraphics[width=7.0cm]{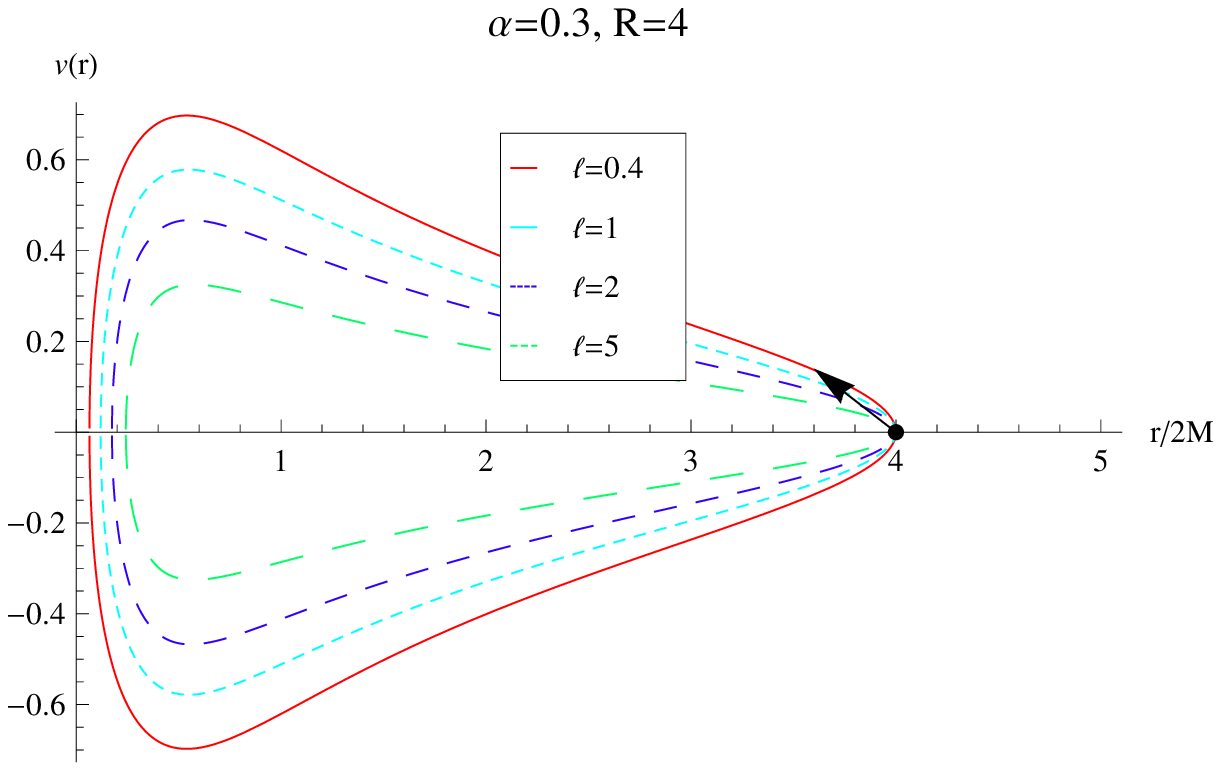}
\caption{The velocity $v(r)$ of an infalling particle for starting at rest at different radii $R$ (Left) and at the radii $R=4$ (Right) with different coupling constant $\ell$ of Einstein-Gauss-Bonnet-bumblebee black hole, where the dot shows the initial position and the arrows are the corresponding direction in which the velocity starts to increase. }
\end{center}
\end{figure}
The left panel of Fig. 3 shows that, if the coupling constant $\ell$ becomes bigger, the short distance which it can reach is farther away from the point $r=0$. Therefore, if $\ell$ is sufficiently large enough\footnote{In Ref.\cite{casana}, the authors assume $\ell\ll1$ and investigate some of its upper bounds, which provides a sensitivity at the $10^{-15}$ level and for future missions at the $10^{-19}$ level. Here we can also assume it is bigger than unit since that at or above Planck scale, Lorentz invariance maybe invalid. }, an infalling particle never reaches the singularity $r=0$. In this sense, we can say that the Einstein-Gauss-Bonnet-bumblebee black hole can be free from the singularity problem.

\subsection{Thermodynamics and phase transition}

From the formula in \cite{ding2008} and with Eq. (\ref{metricm}), this EGB bumblebee black hole's Hawking temperature is
\begin{eqnarray}\label{}
T=\frac{\sqrt{1+\ell}}{4\pi}f'(r_+)=\frac{\sqrt{1+\ell}}{2\pi}\frac{r_+-M}
{(1+\ell)r_+^2+2\alpha },
\end{eqnarray}
which is shown in Fig. 2. It is easy to see that the constant $\ell$ decreases the black hole's Hawking temperature, besides the smaller  mass of the black hole, the lower temperature is.
\begin{figure}[ht]\label{}
\begin{center}
\includegraphics[width=8.0cm]{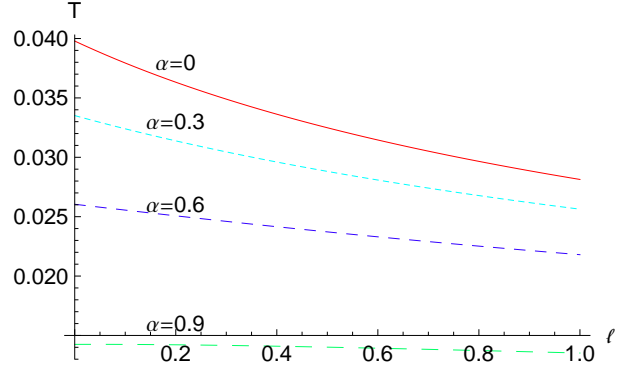}
\caption{Variety of Hawking temperature $T$ with different coupling constant $\ell$ of Einstein-Gauss-Bonnet-bumblebee black hole for
 different expansion parameter $\alpha$. Note that $\alpha=0.3,0.6,0.9$ correspond to the black hole mass $M=M_*/\sqrt{0.3}, M=M_*/\sqrt{0.6}, M=M_*/\sqrt{0.9}$, respectively. }
\end{center}
\end{figure}
As shown in Ref. \cite{wei2020}, the Gauss-Bonnet coupling constant $\alpha$ is treated as a new thermodynamical variable. Then we can construct the black hole first law,
\begin{eqnarray}\label{}
dM=TdS+\mathcal{A}d\alpha.
\end{eqnarray}
$\mathcal{A}d\alpha$ is a work term, $\mathcal{A}$ conjugates to $\alpha$ and it can be called a coupling force. Since $\alpha$ has a dimension of area, it can be termed as a coupling area.
For fixed $\alpha$, the black hole entropy is
\begin{eqnarray}\label{}
S=\int \frac{dM}{T}=\sqrt{1+\ell}\pi r_+^2+\frac{2\pi\alpha}{\sqrt{1+\ell}}\ln\frac{r_+^2}{\alpha}.
\end{eqnarray}
The first term is related to the horizon area $A=4\pi r_+^2$ as $\sqrt{1+\ell}A/4$, which is slightly different from the usual result $S=A/4$. The second term, a logarithmic term, arises as a subleading correction to the Bekenstein-Hawking area formula, which is universal in some quantum theories of gravity \cite{cai}.
If $\ell\rightarrow0$, it can recover the result in Ref. \cite{wei2020}.
The coupling force can be calculated as \cite{wei2020},
\begin{eqnarray}\label{}
\mathcal{A}=\left( \frac{\partial M}{\partial \alpha}\right)_S=\frac{\alpha+(2+\ell)r_+^2+(\alpha-r_+^2)\ln \frac{r_+^2}{\alpha}}{2r_+[2\alpha+(1+\ell)r_+^2]},
\end{eqnarray}
With the above thermodynamical quantities, there exists the following Smarr formula
\begin{eqnarray}\label{}
M=2TS+2\alpha \mathcal{A}.
\end{eqnarray}

Now we focus on the thermodynamical phase structure of EGB bumblebee black hole. The Schwarzschild black hole has only one thermodynamical phase of negative heat capacity, showing that the isolated black hole is thermodynamically unstable \cite{weicpc}. However, after the black hole acquires other hairs, such as charge or spin, another phase of positive heat capacity will appear. At the joined point of these two phases, the heat capacity goes to infinity.
Here, we find that this second order thermodynamic phase transition also can  occur when the Gauss-Bonnet term is considered. The heat capacity at a fixed $\alpha$ is
\begin{eqnarray}\label{}
C_\alpha=T\left( \frac{\partial S}{\partial T}\right)_\alpha=\frac{2\pi}{\sqrt{1+\ell}}\frac{(r_+^2-\alpha)[(1+\ell)r_+^2+2\alpha ]^2}{2\alpha^2+(5+3\ell)\alpha r_+^2-(1+\ell)r_+^4},
\end{eqnarray}
which is shown in Fig. 4 with different black hole mass $M$ and bumblebee coupling constant $\ell$.
\begin{figure}[ht]\label{}
\begin{center}
\includegraphics[width=10.0cm]{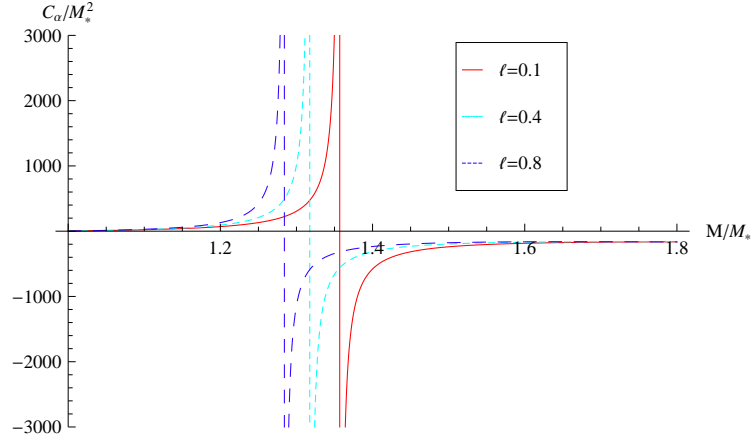}
\caption{Variety of the heat capacity $C_\alpha$ vs. black hole mass $(M>M_*)$  for the EGB bumblebee black hole with different bumblebee coupling constant $\ell$. The black hole mass at the joined point are $M_a=1.36M_*, 1.32M_*, 1.28M_*$,  corresponding to $\ell=0.1,0.4,0.8$, respectively.}
\end{center}
\end{figure}
It is clear that when the black hole mass $M$ is small $(M_*<M<M_a)$, the heat capacity $C_\alpha$ is positive. Meanwhile, for the large mass case, $C_\alpha$ becomes negative, just like a Schwarzschild black hole.  $C_\alpha$ diverges at the joined point
\begin{eqnarray}\label{}
M_a=\left[\frac{-3\ell^2+6\ell+13+(3+\ell)\sqrt{9\ell^2-38\ell+33}}{16(1+\ell)}\right]^{1/2}M_*.
\end{eqnarray}
Therefore, there are two phases for the EGB bumblebee black hole: small stable black hole phase and large unstable black hole phase.

\section{Cosmological solution in Einstein-Gauss-Bonnet-bumblebee gravity}
Many observational data show that the current universe is in a phase of accelerated expansion and in which the approximately 70\% of the total energy content of the universe is dark energy \cite{perlmutter}. The cosmological constant $\Lambda$ is the simplest candidate to describe this dark energy. We show that in this section, the LV bumblebee field and the EGB term can take the place of the cosmological constant in Einstein-Gauss-Bonnet-bumblebee gravity model.

 The Friedmann-Robertson-Walker metric is
\begin{eqnarray}\label{metricc}
&&ds^2=-dt^2+a^2(t)\Big(\frac{dr^2}{1-kr^2}+r^2d\Omega_{D-2}^2\Big),
\end{eqnarray}
where $k=1,0,-1$, corresponding to closed, flat or open cosmos. The function $a(t)$ is the scale factor.
To maintain the assumption of the large-scale homogeneity and isotropy of Universe, the bumblebee field is assumed as following (as in Ref. \cite{capelo})
\begin{eqnarray}
&&B_\mu=(B(t),0,\cdots,0),
\end{eqnarray}
where there is no spatial orientation.
Then the field strength tensor vanishes, $B_{\mu\nu}=0$, and the bumblebee motion equation (\ref{motion}) becomes
\begin{eqnarray}
&&2V' B_\nu=\frac{\varrho}{\kappa}B^\mu R_{\mu\nu}.
\end{eqnarray}
The tensor $T^{B}_{\mu\nu}$ becomes
\begin{eqnarray}\label{momentumB}
&&T_{\mu\nu}^B=- g_{\mu\nu}V+\frac{\varrho}{\kappa}\tilde{B}_{\mu\nu},
\end{eqnarray}
where $\tilde{B}_{\mu\nu}$ is
\begin{eqnarray}
&&\tilde{B}_{\mu\nu}=\frac{1}{2}g_{\mu\nu}B^{\alpha}B^{\beta}R_{\alpha\beta}
-B_{\mu}B^{\alpha}R_{\alpha\nu}+\frac{1}{2}\nabla_{\alpha}\nabla_{\mu}(B^{\alpha}B_{\nu})
\nonumber\\
&&\qquad\qquad
+\frac{1}{2}\nabla_{\alpha}\nabla_{\nu}(B^{\alpha}B_{\mu})
-\frac{1}{2}\nabla^2(B_{\mu}B_{\nu})-\frac{1}{2}
g_{\mu\nu}\nabla_{\alpha}\nabla_{\beta}(B^{\alpha}B^{\beta}).
\end{eqnarray}
Supposing that the matter contribution is equivalent to an ideal fluid $T^M_{\mu\nu}=diag(\rho,p,\cdots,p)$ with the energy density of matter $\rho$ and its pressure $p$, then the gravitational equations of motion are,
\begin{eqnarray}\label{einstein2}
G_{\mu\nu}=-\kappa g_{\mu\nu}V+\kappa T^M_{\mu\nu}+\varrho\tilde{B}_{\mu\nu}+\frac{2\alpha}{D-4}\kappa T^{GB}_{\mu\nu}.
\end{eqnarray}
The non-zero components of tensors $G_{\mu\nu},\tilde{B}_{\mu\nu}$ and $T^{GB}_{\mu\nu}$ are also shown in the appendix. The $tt$ and $rr$ components of Eq. (\ref{einstein2}) are
\begin{eqnarray}
&&\frac{1}{2}(D-1)(D-2)\Big[\frac{k}{a^2}+H^2+2\alpha\kappa(D-3)\big(\frac{k}{a^2}+H^2\big)^2-\varrho B^2H^2\Big]=\kappa (\rho+V)+(D-1)\varrho HB\dot B,\label{Cos-tt}\\
&&(D-2)\frac{\ddot a}{a}\Big[4\alpha\kappa(D-3)(\frac{k}{a^2}+H^2)+1-\varrho B^2\Big]+\frac{1}{2}(D-2)(D-3)\Big[\frac{k}{a^2}+H^2(1-\varrho B^2)+2\alpha\kappa(D-5)\big(\frac{k}{a^2}+H^2\big)^2\Big]\nonumber\\&&
\qquad=\kappa (-p+V)+\varrho\Big[2(D-2) HB\dot B+\dot B^2+B\ddot B\Big]\label{Cos-rr}.
\end{eqnarray}
Eq. (\ref{Cos-rr}) can be rewritten, with $\dot H=\ddot a/a-H^2$ and Eq. (\ref{Cos-tt}), as following,
\begin{eqnarray}
&&(D-2)\dot H\Big[4\alpha\kappa(D-3)(\frac{k}{a^2}+H^2)+1-\varrho B^2\Big]-(D-2)\frac{k}{a^2}\Big[1+4\alpha\kappa(D-3)\big(\frac{k}{a^2}+H^2\big)\Big]\nonumber\\&&
\qquad=-\kappa (\rho+p)+\varrho\Big[(D-3) HB\dot B+\dot B^2+B\ddot B\Big]\label{Cos-rr2}.
\end{eqnarray}

When $D\rightarrow4$, Eqs. (\ref{Cos-tt}) and (\ref{Cos-rr2}) are
\begin{eqnarray}\label{costt}
&&3\Big[\frac{k}{a^2}+H^2+2\alpha\kappa(\frac{k}{a^2}+H^2)^2-\varrho B^2H^2\Big]=\kappa (\rho+V)+3\varrho HB\dot B,\\
&&2\dot H\Big[4\alpha\kappa(\frac{k}{a^2}+H^2)+1-\varrho B^2\Big]-\frac{2k}{a^2}\Big[1+4\alpha\kappa(\frac{k}{a^2}+H^2)\Big]=-\kappa (\rho+p)+\varrho\Big[HB\dot B+\dot B^2+B\ddot B\Big]\label{cosrr},
\end{eqnarray}
where $\kappa=8\pi G$ in four dimensions.
If one lets $k=0,\varrho=V=0$, the above both equations can reduce to those in Ref. \cite{lin2020}; if one lets $\alpha=k=0,p=0$, they are the same as those in Ref. \cite{capelo}. The conservation of energy equation is
\begin{eqnarray}\label{energy0}
&&\dot\rho=-3H(p+\rho)-\frac{3\varrho}{\kappa}\frac{\ddot{a}}{a}B(HB+\dot B)+\frac{3\varrho}{\kappa}\frac{\dddot{a}}{a}B^2,
\end{eqnarray}
which is interesting that it doesn't relate to expansion parameter $\alpha$.

If $k=0$ and $\dot B=\ddot B=0$, i.e., the bumblebee field $B(t)$ doesn't evolve with time and its potential rest at a constant $V=V_0$ and $B=B_0$, the Eqs. (\ref{costt}), (\ref{cosrr}) and (\ref{energy0}) become
\begin{eqnarray}\label{}
&&3(1-\varrho B_0^2)H^2+6\alpha\kappa H^4=\kappa (\rho+V_0),\label{dens}\\
&&2\dot H(4\alpha\kappa H^2+1-\varrho B_0^2)=-\kappa (\rho+p),\\
&&\dot \rho=-3H(\rho+p)+\frac{3\varrho}{\kappa}B^2(\ddot H+2H\dot H).
\end{eqnarray}
Consider an empty space case $\rho=p=0$, then one can obtain a de Sitter universe solution $\dot H=0$ and
\begin{eqnarray}\label{}
a(t)=a_0e^{H_0(t-t_0)},
\end{eqnarray}
with
\begin{eqnarray}\label{hubb}
&&H_0^2=\frac{1}{4\alpha\kappa}\Big[\sqrt{\big(1-\varrho B_0^2\big)^2+8\alpha\kappa^2 V_0/3}-(1-\varrho B_0^2)\Big],\qquad \dot\rho=0,
\end{eqnarray}
where $a_0,\;t_0,\;H_0$ are scale factor, time and Hubble constant in the present epoch, respectively.
If the constant $\alpha$ is a very small quantity, then
\begin{eqnarray}\label{}
&&H_0^2=\frac{\kappa V_0}{3(1-\varrho B_0^2)},
\end{eqnarray}
which is the same as that in Ref. \cite{capelo}, corresponding to a cosmological constant $\Lambda=\kappa V_0/(1-\varrho B_0^2)$, an
universe without matter that is dominated by the bumblebee field alone. The de-Sitter universe cannot be taken as the model of our realistic universe. Because the authors in Ref. \cite{perlmutter} found that in the flat universe, the density of matter is about $\Omega_M=0.28^{+0.09}_{-0.08}$, and $1=\Omega_M+\Omega_\Lambda$, where $\Omega_\Lambda$ is the density of dark energy from the cosmological constant $\Lambda$. But at late times, the evolution will be dominated essentially by the cosmological constant---dark energy \cite{pad}.

We can define the following effective density parameters,
 \begin{eqnarray}\label{}
&&\Omega_V\equiv \frac{\kappa V_0}{3H_0^2},\;\Omega_B\equiv \varrho B_0^2,\;\Omega_\alpha\equiv 2\alpha\kappa H_0^2,
\end{eqnarray}
corresponding to the effective density of the bumblebee potential $V_0$, the bumblebee field $B_0$ and $\alpha$, respectively. Then Eq. (\ref{dens}) can be rewritten as
 \begin{eqnarray}\label{density}
&&1=\Omega_V+\Omega_B-\Omega_\alpha,
\end{eqnarray}
when the matter density $\rho=0$.
From the density Eq. (\ref{density}), it is easy to see that in the de-Sitter universe, corresponding to dark energy dominated era of the universe, both the bumblebee field and the Gauss-Bonnet term have contributions to the expansion of the universe and actually act as a form of dark energy.

\section{Summary and discussion}

In this paper, we have studied Einstein-Gauss-Bonnet(EGB) gravity coupled to a bumblebee field. We obtain an exactly black hole solution and cosmological solutions in four dimensional spacetime by a regularization scheme. The bumblebee field doesn't affect the locations of the black hole horizon. This black hole is different from Schwarzschild black hole due to that the gravitational potential has a minimum negative value in the black hole interior.  And it is also different from the 4D EGB black hole \cite{lin2020} and EGB Born-Infeld black hole \cite{yang}, due to that it has a positive value $1+\ell$ at short distance $r\rightarrow0$. Other effects of bumblebee fields on the gravitational potential are, decreasing the minimum of the gravitational potential in the black hole interior and increasing the positive finite value at short distance.

This positive gravitational potential $1+\ell$ provides a repulsive force at short distance to an infalling particle. If the coupling constant $\ell$ is very much big, then this positive gravitational potential is sufficiently large to prevent an infalling particle approaching the origin, i.e., the infalling particle never hits the singularity $r=0$. In this sense, this black hole can be free from the singularity problem.

The black hole entropy includes two terms: slightly modified area term and subleading correction logarithmic term.  The black hole first law and Smarr formula can both be constructed. There exist two phases:  small stable black hole phase and large unstable black hole phase.

In the cosmological context, it is interesting that the Gauss-Bonnet term hasn't any effect on the conservation of energy equation. The obtained cosmological solutions show that, in an universe with no matter content, there exist a de Sitter Universe solution. In this case, both the bumblebee field and the Gauss-Bonnet term have contributions to the present Hubble constant $H_0$ and act as a dark energy component driving Universe accelerated expansion. The effects of the Gauss-Bonnet term on the bumblebee potential $V_0$ are: increasing it if  $\alpha>0$; decreasing it if $\alpha<0$.

This Einstein-Gauss-Bonnet theory is proved by Oikonomou {\it et al} that it can be rendered viable and compatible with the GW170817 event \cite{oikonomou}, which shows that the gamma rays arrived almost simultaneously with the gravitational waves \cite{abbott}.  In their works, the matter field $\mathcal{L}_M=-\partial_\mu\phi \partial^\mu\phi/2-V(\phi)$ and the coupling constant $2\alpha/(D-4)$ is replaced by a scalar coupling function $-\xi(\phi)/2$ in the action (\ref{action}) when the bumblebee field absents, where $\phi$ is a scalar field. They then considered the cosmological tensor perturbation speed and found that it is compatible with the Planck 2018 data \cite{akrami}. Is it also compatible with the GW170817 event when the bumblebee field involves? We will consider this issue in the near future.

Note that one can obtain a $D$-dimensional Schwarzschild-like solution in bumblebee gravity. In the first branch of the metric function (\ref{dmetric}), one lets $\alpha\rightarrow0$ and uses (\ref{dell}), then can get that
\begin{eqnarray}\label{}
ds^2=-\Big(1-\frac{2GM}{r^{D-3}}\Big)dt^2
+\frac{1+\ell}{(1-\frac{2GM}{r^{D-3}})}dr^2+r^2d\Omega_{D-2}^2.
\end{eqnarray}

\begin{acknowledgments} The authors would like to thank Prof. Ralf Lehnert(Indiana University) for the helpful suggestion. This work was supported by the Scientific Research Fund of the Hunan Provincial Education Department under No. 19A257, the National Natural Science Foundation (NNSFC)
of China (grant No. 11247013), Hunan Provincial Natural Science Foundation of China (grant No. 2015JJ2085 and No. 2020JJ4284).
\end{acknowledgments}
\appendix\section{Some quantities I}
In this appendix, we showed the nonezero components of Einstein's tensor for the metric (\ref{metric}). They are as following (cf. \cite{wiltshire})
\begin{eqnarray}
&&G_{00}=\frac{(D-2)e^{2\phi-2\psi}}{2r^2}\Big[(D-3)(e^{2\psi}-1)+2r\psi'\Big],\\
&&G_{11}=\frac{(D-2)}{2r^2}\Big[(D-3)(1-e^{2\psi})+2r\phi'\Big],\\
&&G_{22}=e^{-2\psi}\Big[\frac{(D-3)(D-4)}{2}(1-e^{2\psi})+(D-3)r(\phi'-\psi')
+r^2(\phi''+\phi'^2-\phi'\psi')\Big],\\
&&G_{ii}=G_{22}\prod^{i-2}_{j=1}\sin^2\theta_j,\\
&&R_{11}=\frac{(D-2)}{r}\psi'-(\phi''+\phi'^2-\phi'\psi'),
\end{eqnarray}
where $(D-1)\geq i\geq3$. $\bar B_{\mu\nu}$  are
\begin{eqnarray}
&&\bar B_{00}=\frac{\varrho b^2e^{2\phi-2\psi}}{2r^2}\Big[(D-2)(D-3)-(D-2)r\psi'-r^2(\phi''+\phi'^2
-\phi'\psi')\Big],\\
&&\bar B_{11}=-\frac{\varrho b^2}{2r^2}\Big[(D-2)(D-3)-(D-2)r(\psi'-2\phi')+r^2(\phi''+\phi'^2-\phi'\psi')\Big],\\
&&\bar B_{22}=-\frac{\varrho b^2e^{-2\psi}}{2}\Big[(D-3)(D-4)-(D-4)r\psi'+2(D-3)r\phi'+r^2(\phi''
+\phi'^2-\phi'\psi')\Big],\\
&&\bar B_{ii}=\bar B_{22}\prod^{i-2}_{j=1}\sin^2\theta_j.
\end{eqnarray}
The nonzero components of the Gauss-Bonnet term $T^{GB}_{\mu\nu}$ are
\begin{eqnarray}
&&T^{GB}_{00}=\frac{(D-2)(D-3)(D-4)}{r^2}e^{2\phi-2\psi}
\Big[\frac{2}{r}(1-e^{-2\psi})\psi'+\frac{(D-5)}{r^2}(\frac{e^{2\psi}+e^{-2\psi}}{2}-1)\Big],\\
&&T^{GB}_{11}=\frac{(D-2)(D-3)(D-4)}{r^2}
\Big[-\frac{2}{r}(1-e^{-2\psi})\phi'+\frac{(D-5)}{r^2}(\frac{e^{2\psi}+e^{-2\psi}}{2}-1)\Big],\\
&&T^{GB}_{22}=(D-3)(D-4)e^{-2\psi}\Big\{-4e^{-2\psi}\phi'\psi'+(1-e^{-2\psi})
\Big[\frac{(D-5)(D-6)}{2r^2}(e^{2\psi}-1)\nonumber\\&&\qquad\qquad
-\frac{2(D-5)}{r}(\phi'-\psi')
-2(\phi''+\phi'^2-\phi'\psi')\Big]\Big\},\\
&&T^{GB}_{ii}=T^{GB}_{22}\prod^{i-2}_{j=1}\sin^2\theta_j.
\end{eqnarray}
\section{Some quantities II}
We showed the nonezero components of Einstein's tensor for the metric (\ref{metricc}). They are as following (cf. \cite{pavluchenko})
\begin{eqnarray}
&&G_{00}=\frac{(D-1)(D-2)}{2}\frac{k+\dot{a}^2}{a^2},\\
&&G_{11}=-\frac{(D-2)}{1-kr^2}\Big[a\ddot{a}+\frac{D-3}{2}(k+\dot{a}^2)\Big].
\end{eqnarray}
 $\tilde B_{\mu\nu}$  are
\begin{eqnarray}
&&\tilde B_{00}=(D-1)\frac{B\dot{a}}{a^2}\Big[a\dot{B}+\frac{(D-2)}{2}B\dot{a}\Big],\\
&&\tilde B_{11}=-\frac{1}{1-kr^2}\Big[\frac{(D-2)(D-3)}{2}B^2\dot{a}^2+(D-2)Ba(2\dot{B}\dot{a}+B\ddot{a})
+a^2(\dot{B}^2+B\ddot{B})\Big].
\end{eqnarray}
The nonzero components of the Gauss-Bonnet term $T^{GB}_{\mu\nu}$ are
\begin{eqnarray}
&&T^{GB}_{00}=-(D-1)(D-2)(D-3)(D-4)\frac{(k+\dot{a}^2)^2}{2a^4},\\
&&T^{GB}_{11}=(D-2)(D-3)(D-4)\frac{k+\dot{a}^2}{1-kr^2}
\Big[2\frac{\ddot{a}}{a}+\frac{(D-5)(k+\dot{a}^2)}{2a^2}\Big].
\end{eqnarray}


\vspace*{0.2cm}
 \begin{thebibliography}{99}
 \baselineskip=0.6 cm


\bibitem{pavluchenko}S. A. Pavluchenko, Phys. Rev. D {\bf94}, 024046 (2016).
\bibitem{wiltshire}D. L. Willtsire, Phys. Lett. B {\bf169}, 36 (1986).
\bibitem{lovelock} D. Lovelock, J. Math. Phys. {\bf12}, 498 (1971); {\it ibid} {\bf13}, 874 (1972).
\bibitem{zwiebach} B. Zwiebach, Phys. Lett. B {\bf156}, 315 (1985).
\bibitem{lin2020} D. Glavan and Chunshan Lin, Phys. Rev. Lett. {\bf 124}, 081301 (2020).
\bibitem{shu} F. Shu, Phys. Lett. B {\bf811}, 135907 (2020); W. Ai, 
 Commun. Theor. Phys. {\bf72}, 095402 (2020);  M. Gurses, T. C. Sisman and B. Tekin, Eur. Phys. J. C {\bf80}, 647 (2020);
  S. Mahapatra, Eur. Phys. J. C {\bf80}, 992 (2020);
   M. Gurses, T. C. Sisman and B. Tekin, Phys. Rev. Lett. {\bf125}, 149001 (2020);
  J. Arrechea, A. Delhom and A. Jim\'{e}nez-Cano,  Phys. Rev. Lett. {\bf125}, 149002 (2020).
\bibitem{arrechea}  J. Arrechea, A. Delhom and A. Jim\'{e}nez-Cano,  Chin. Phys. C {\bf45} 013107 (2021).

\bibitem{hennigar}R. A. Hennigar, D. Kubiznak, R. B. Mann and C. Pollack, 
 J. High Energy Phys. {\bf07}, 027 (2020); A. Casalino, A. Colleaux, M. Rinaldi and S. Vicentini,
 Phys. Dark Universe, {\bf31}, 100770 (2021); H. Lu and Y. Pang, 
 Phys. Lett. B {\bf809}, 135717 (2020); T. Kobayashi, 
J. Cos. Astro. Phys. {\bf07}, 013  (2020).
\bibitem{fernandes}P. G. S. Fernandes, Phys. Lett. B {\bf805}, 135468 (2020).
\bibitem{yang} Ke Yang, Bao-Min Gu, Shao-Wen Wei and Yu-Xiao Liu, Eur. Phys. J. C {\bf80}, 662 (2020).
\bibitem{feng}Jia-Xi Feng, Bao-Min Gu and Fu-Wen Shu, Phys. Rev. D {\bf103}, 064002 (2021).
\bibitem{kumar2004} A. Kumar, S. G. Ghosh, 
    arXiv:2004.01131; A. Kumar, R. Kumar, 
arXiv:2003.13104.
\bibitem{wei2021}S.-W. Wei, Y.-X. Liu, 
 Eur. Phys. J. Plus {\bf136}, 436 (2021); R. Kumar, S. G. Ghosh, 
 JCAP {\bf07}, 053 (2020). 
\bibitem{ghosh} S. G. Ghosh, R. Kumar, 
    Class. Quant. Grav. {\bf37},  245008 (2020); 
S. G. Ghosh, S. D. Maharaj, 
 Phys. Dark Univ. {\bf30}, 100687 (2020); 
S. G. Ghosh, S. D. Maharaj, 
Phys. Dark Univ. {\bf31}, 100793  (2021). 

\bibitem{kostelecky198939} V. A. Kosteleck\'{y} and S. Samuel, Phys. Rev. D {\bf39}, 683 (1989).
\bibitem{mattingly} D.~Mattingly,
Living Rev.\ Rel.\  {\bf 8}, 5 (2005).

\bibitem{carroll}S. M. Carroll, J. A. Harvey, V. A. Kostelecky, C. D. Lane,
and T. Okamoto, Phys. Rev. Lett. {\bf87}, 141601 (2001); I.
Mocioiu, M. Pospelov, and R. Roiban, Phys. Lett. B {\bf489},
390 (2000); A. F. Ferrari, M. Gomes, J. R. Nascimento, E.
Passos, A. Yu. Petrov, and A. J. da Silva, Phys. Lett. B {\bf652},
174 (2007).
\bibitem{bertolami69}O. Bertolami, R. Lehnert, R. Potting, and A. Ribeiro,
Phys. Rev. D {\bf69}, 083513 (2004); V. A. Kostelecky\'{y}, R.
Lehnert, and M. Perry, Phys. Rev. D {\bf68}, 123511 (2003);
R. Jackiw and S.-Y. Pi, Phys. Rev. D {\bf68}, 104012
(2003).
\bibitem{gambini} R. Gambini and J. Pullin, Phys. Rev. D {\bf59}, 124021 (1999);
J. R. Ellis, N. E. Mavromatos, and D. V. Nanopoulos, Gen.
Relativ. Gravit. {\bf32}, 127 (2000).
\bibitem{burgess03} C. P. Burgess, J. Cline, E. Filotas, J.
Matias, and G.D. Moore, J. High Energy Phys. {\bf03} (2002)
043; A. R. Frey, J. High Energy Phys. {\bf04} (2003) 012; J.
Cline and L. Valc\'{a}rcel, J. High Energy Phys. {\bf03} (2004)
032.
\bibitem{fernando} A. Fernando, T. Clark, Gen. Relativ. Grav. {\bf46}, 1834 (2014).
\bibitem{jacobson} T. Jacobson and D. Mattingly, Phys. Rev. D {\bf64}, 024028
(2001); T. Jacobson, Proc. Sci. QG-PH (2007) 020.

\bibitem{kostelecky2004} V. A. Kosteleck\'{y}, Phys. Rev. D {\bf69}, 105009 (2004).
\bibitem{bluhm125007}R. Bluhm, N. L. Gagne, R. Potting and A. Vrublevskis, Phys. Rev. D {\bf77}, 125007 (2008).


\bibitem{dickinson} M. H. Dickinson, F. O. Lehmann and S. P. Sane, Science {\bf284}, 1954 (1999).
\bibitem{kostelecky198940} V. A. Kosteleck\'{y} and S. Samuel, Phys. Rev. D {\bf40}, 1886 (1989).
\bibitem{casana} R. Casana and A. Cavalcante, Phys. Rev. D {\bf97}, 104001 (2018).
\bibitem{ding2020} C. Ding, C. Liu, R. Casana and A. Cavalcate, Eur. Phys. C {\bf80}, 178 (2020).
\bibitem{li}Z. Li and A. \"{O}vg\"{u}n, Phys. Rev. D {\bf101}, 024040 (2020).
\bibitem{jha} S. K. Jha and A. Rahaman, arXiv: 2011.14916, (2021).
\bibitem{ding2021} C. Ding and X. Chen, Chin. Phys. C {\bf45}, 025106 (2021).
\bibitem{bluhm} R. Bluhm, Shu-Hong Fung and V. A. Kosteleck\'{y}, Phys. Rev. D {\bf77}, 065020
(2008).
\bibitem{bluhm2005} R. Bluhm and V. A. Kosteleck\'{y}, Phys. Rev. D {\bf71}, 065008
(2005).
\bibitem{bertolami} O. Bertolami and J. P\'{a}ramos, Phys. Rev. D {\bf72}, 044001
(2005).

\bibitem{boulware} D. G. Boulware and S. Deser, Phys. Rev. Lett. {\bf55}, 2656 (1985).
\bibitem{chandrasekhar} S. Chandrasekhar, \textit{The mathematical theory of black holes,
Handbook of Mathematical}, Clarendon press, Oxford, (1985).
\bibitem{ding2008} C. Ding and J. Jing, Class. Quantum Grav. {\bf25}, 145015 (2008).
\bibitem{cai} R.-G. Cai, L.-M. Cao and N. Ohta, 
    JHEP {\bf04}, 082 (2010). 

\bibitem{wei2020} Shao-Wen Wei and Yu-Xiao Liu,  Phys. Rev. D {\bf101}, 104018 (2020).
\bibitem{weicpc} Shao-Wen Wei and Yu-Xiao Liu, Chin. Phys. C {\bf44}, 115103 (2020).
\bibitem{perlmutter}S. Perlmutter {\it et al}, Astrophys. J. {\bf517} 565 (1999).
\bibitem{capelo} D. Capelo, Phys. Rev. D {\bf91}, 104007 (2015).

\bibitem{pad} T. Padmanabhan, \textit{Gravitation: Foundations and Frontiers}, Cambridge University Press, Londan, (2010).
\bibitem{oikonomou}V. K. Oikonomou, Class. Quantum Grav. {\bf38}, 195025 (2021); S. D. Odintsov, V. K. Oikonomou and F. P. Fronimos, Nucl. Phys. B {\bf958}, 115135 (2020); S. D. Odintsov and V. K. Oikonomou, Phys. Lett. B {\bf805}, 135437 (2020); V. K. Oikonomou and F. P. Fronimos, Europhys. Lett. {\bf131}, 30001 (2020).
\bibitem{abbott}     B. P. Abbott {\it et al}.
     Astrophys. J. {\bf848}, L12 (2017).
\bibitem{akrami}  Y. Akrami, {\it et al.}, Planck Collaboration, arXiv: 1807.06211.
\end{thebibliography}
\end{document}